\documentstyle[epsfig]{europhys}

\def\stars{\bigskip\centerline{***}\medskip}

\newif\ifboo \boofalse

\def\Review#1{\boofalse{\it #1},}
\def\Name#1{{\sc #1},}
\def\Vol#1{\ifboo Vol. {\bf #1}\else{\bf #1}\fi}
\def\Year#1{\ifboo #1\else(#1)\fi}
\def\Book#1{\bootrue{\it #1},}
\def\Page#1{\ifboo {\rm p. #1}\else{\rm #1}\fi}

\def\bra#1{\langle \, {#1} \, | \;}
\def\ket#1{\; | \, {#1} \, \rangle}
\newcommand{\braket}[2]{\langle \, {#1} \, | \, {#2} \, \rangle}
\newcommand{\op}[1]{%
    \fontdimen12\textfont3=2pt\fontdimen12\scriptfont3=1.4pt%
    \!\null\mathop{\vphantom{#1}\smash{#1}}\limits_{\sim}\null\!}
\newcommand{\EinsOp}
           {\;\smash{\raisebox{-1.1ex}{$\!\!\stackrel{\!\mbox{1}
            \hspace{-0.4ex}\rule[0.0ex]{0.06ex}{1.60ex}}{\sim}$}}}
\newcommand{\dint}{\mbox{d}}
\newcommand{\Mean}[1]{\big\langle\big\langle \; {#1}\; 
            \big\rangle\big\rangle}
\def\half{{\frac{1}{2}}\;}
\newcommand{\fmref}[1]{(\protect\ref{#1})}
\newcommand{\xref}[1]{\protect\ref{#1}}
\renewcommand{\Re}[1]{\mbox{Re}\left(#1\right)\;}
\renewcommand{\Im}[1]{\mbox{Im}\left(#1\right)\;}

\begin{document}
\euro{??}{?}{1-6}{1998}
\Date{August 1998}
\shorttitle{J.SCHNACK THERMODYNAMICS WITH COHERENT STATES}
\title{Thermodynamics of the harmonic oscillator using coherent states}
\author{J. Schnack\footnote{Email: jschnack\char'100uos.de,
WWW:~http://obelix.physik.uni-osnabrueck.de/$\sim$schnack}}
\institute{Universit\"at Osnabr\"uck, Fachbereich Physik \\ 
         Barbarastr. 7, D-49069 Osnabr\"uck}
%
%
\rec{}{}
%
%
%
\pacs{
\Pacs{05}{30.-d}{Quantum statistical mechanics}
\Pacs{05}{30.Ch}{Quantum ensemble theory}
\Pacs{02}{70.Ns}{Molecular dynamics and particle methods}
      }
\maketitle
\begin{abstract}
The ongoing discussion whether thermodynamic properties can be
extracted from a (possibly approximate) quantum mechanical time
evolution using time averages is fed with an instructive
example. It is shown for the harmonic oscillator how the Hilbert
space or an appropriately defined phase space must be populated
in terms of coherent states in order to obtain the quantum
result respectively the classical one.
\end{abstract}

\section{Introduction}

Transport models which employ wave packet dynamics are widely
used in atomic, solid state and nuclear physics today. Examples
range from the description of hydrogen plasmas \cite{EbM97} over
the investigation of critical properties of noble gases
\cite{OhR97B} to simulations of heavy ion collisions in nuclear
physics \cite{AiS86,Fel90,OHM92}. In all of these models the
quantum mechanical time evolution is approximated by the
evolution of a (possibly antisymmetrized) product state composed
of Gaussian single-particle wave packets (squeezed coherent
states \cite{KlS85})
\begin{eqnarray}
\braket{\vec{x}}{\vec{r},\vec{p},a}
&\propto&
\exp\left\{-\frac{(\,\vec{x}-\vec{r}\,)^2}{2\,a}
+\frac{i}{\hbar}\vec{p}\cdot\vec{x}\right\} 
\ ,
\end{eqnarray}
where the time dependence of the state is given through the time
dependence of the parameters, as there are the mean position
$\vec{r}$, the mean momentum $\vec{p}$ and in some cases the
complex width $a$.

Although these models were designed to describe non-equilibrium
situations their thermodynamic equilibrium properties are
important for two reasons. Firstly, their statistical properties
have great influence on final observables like fragment
distributions in collisions. Secondly, long time evolutions are
used to determine equilibrium behavior and to estimate quantities
like the degree of ionization of a plasma \cite{EbM97} or the
caloric curve for the nuclear liquid-gas phase transition
\cite{ScF97}. The last point is of special interest since
approximate quantum evolutions are a possible way to examine
thermodynamics properties in cases where the energy spectrum and
thus the partition function is not known. In these cases the
time-dependent Schr\"odinger equation cannot be solved, too.

Whereas thermostated time evolutions are on firm grounds in
classical mechanics \cite{Nos84,Hoo85,KBB90,Nos91} and have been
used successfully as well for equilibrium, e.g. to investigate
classical spin systems, as for non-equilibrium, e.g. to study
glass transitions, in quantum mechanics less
attempts were made, some of them are given in
refs. \cite{EbM97,ScF97,Kus93,OhR93,KTR94A,BDH94,ScF96,OnH96,OhR97,Sch98}.
Following these articles one realizes that the matter is still
under debate.  The questions are twofold. The first concerns the
problem whether it is in principle possible to extract
thermodynamic properties from a quantum mechanical time
evolution. The second question asks in how far a coherent state
used as a trial state in an approximate dynamics like a quantum
molecular dynamics is able to visit the Hilbert space in time
with the correct weight.

Some of the arguments, which doubt a success of time averaging
are loosely speaking like (see for instance
\cite{EbM97,OhR93,KTR94A,KTR94B}): {\it Since the equations of
motion, either approximative ones like in Antisymmetrized
Molecular Dynamics (AMD) \cite{OHM92,OnH96} or Fermionic
Molecular Dynamics (FMD) \cite{Fel90,FeS97} or the Schr\"odinger
equation itself, have a symplectic structure, like Hamilton's
equation of motion in classical mechanics, a time averaging
should lead to classical statistics at its best.  The dynamics
does not know about the discrete structure of the energy levels,
it is not quantized. One can excite a system, which is described
by coherent states, by any small energy, so it does not know
about the finite spacing of energy levels. Coherent states are
quasi classical states, the dynamics must lead to classical
statistics. Or last but not least, the restriction to product
states is equivalent to a mean field calculation, the result of
time averaging is therefore classical.}  This ongoing discussion
got new momentum during the program INT-98-2 \cite{INT98} at the
University of Washington in summer 1998.

It is the aim of this letter to demonstrate with the algebraicly
simple, but instructive example of the harmonic oscillator how
the Hilbert space or an appropriately defined phase space must
be populated in terms of coherent states in order to obtain the
quantum result or the classical one. That means, that one cannot
{\it a priori} say, that a (possibly approximate) quantum
mechanical time evolution is not able to reproduce the correct
ensemble average nor that the restriction of the trial state to
wave packets or products of them impedes correct time averages.

The harmonic oscillator is not a mere example, nowadays it is of
special interest for the dynamics of bosonic or fermionic atoms
contained in magnetic traps \cite{AEM95,DMA95,BSH97} as well as
for all systems that show an equidistant level spacing close to
the ground state like nuclei or Luttinger liquids.

\section{Thermodynamic mean}

Given the Hamilton operator $\op{H}$ of the harmonic oscillator
\begin{eqnarray}
\label{E-1}
\op{H}
=
\hbar\omega\; \left( \op{{a}}^\dagger \op{{a}} 
+ \frac{1}{2} \right)
\ ,
\end{eqnarray}
coherent states are defined as eigenstates of the destruction
operator $\op{{a}}$
\begin{eqnarray}
\label{E-2}
\op{{a}}\; \ket{{z}}
=
{z} \; \ket{{z}}
\ ,
\quad
z = 
\sqrt{\frac{m \omega}{2 \hbar}}\; r
+ \frac{i}{\sqrt{2 m \hbar \omega}}\; p
\ .
\end{eqnarray}
Each coherent state is characterized by a complex parameter $z$
which corresponds to a pair of real parameters $(r,p)$. Coherent
states span the Hilbert space, they are over-complete and obey
the completeness relation \cite{KlS85}
\begin{eqnarray}
\label{E-3}
\EinsOp^{(1)}
&=& \int \frac{\dint^2z}{\pi}\;
\ket{{z}}\bra{{z}}
\ , \quad \dint^2z=\dint\,\Re{z}\dint\,\Im{z}
\\
&=& 
\int \frac{\mbox{d}r\,\mbox{d}p}{(2\pi\hbar)}\;
\ket{{r},\;{p}}\bra{{r},\;{p}}
\nonumber
\ .
\end{eqnarray}
A thermodynamic mean of an observable $\op{B}$ in the
canonical ensemble is given by the trace of this operator
together with the statistical operator. This trace can be
expressed in any basis, it needs not to be the eigenbasis of
$\op{H}$, it may be a basis like those of coherent states
which are characterized by a continuous parameter
\cite{FeS97,Sch96}
\begin{eqnarray}
\label{E-4}
\Mean{\op{B}}
&=& 
\frac{1}{Z(\beta)}
\int \frac{\dint^2z}{\pi}\;
\bra{z}\op{B} e^{-\beta\op{H}} \ket{z}
\\
&=& 
\frac{1}{Z(\beta)}
\int \frac{\dint^2z}{\pi}\;
e^{-\half\beta\hbar\omega}\;
e^{-|z|^2\left(1-e^{-\beta\hbar\omega}\right)}\;
\bra{e^{-\half\beta\hbar\omega} z}\op{B}
\ket{e^{-\half\beta\hbar\omega} z}
\nonumber
\\
&=& 
\frac{1}{Z(\beta)}
\int \frac{\dint^2z}{\pi}\;
e^{\half\beta\hbar\omega}\;
e^{-|z|^2\left(e^{\beta\hbar\omega}-1\right)}\;
\bra{z}\op{B} \ket{z}
\nonumber
\ .
\end{eqnarray}
The partition function then reads
\begin{eqnarray}
\label{E-5}
Z(\beta)
&=& 
\int \frac{\dint^2z}{\pi}\;
\bra{z}e^{-\beta\op{H}} \ket{z}
\\
&=& 
\int \frac{\dint^2z}{\pi}\;
e^{\half\beta\hbar\omega}\;
e^{-|z|^2\left(e^{\beta\hbar\omega}-1\right)}
\nonumber
\ .
\end{eqnarray}
Given the excitation energy
\begin{eqnarray}
\label{E-6}
{\mathcal H}(z)
&=&
\bra{z}\op{H} \ket{z}-\half\hbar\omega 
= 
\hbar\omega |z|^2
= 
\frac{p^2}{2 m} + \half m \omega^2 r^2
\ ,
\end{eqnarray}
the weight with which a coherent state $\ket{z}=\ket{r,p}$
contributes to a thermodynamic mean is just
\begin{eqnarray}
\label{E-7}
w_{\mbox{qm}}(\beta)
&=&
e^{-|z|^2
\left(e^{\beta\hbar\omega}-1\right)}
=
e^{-{\mathcal H}(z)
\left(e^{\beta\hbar\omega}-1\right)/(\hbar\omega)}
\end{eqnarray}
where all terms depending only on the zero point energy have
been omitted since they cancel with the respective terms in the
partition function.

One can now interpret the space of parameters $z=(r,p)$ as a
phase space and reformulate eqs. \fmref{E-4} and \fmref{E-5}
\begin{eqnarray}
\label{E-8}
\Mean{\op{B}}
&=& 
\frac{1}{Z(\beta)}
\int \frac{\mbox{d}r\,\mbox{d}p}{(2\pi\hbar)}\;
w_{\mbox{qm}}(\beta)\;
{\mathcal B}(r,p)
\\
Z(\beta)
&=& 
\int \frac{\mbox{d}r\,\mbox{d}p}{(2\pi\hbar)}\;
w_{\mbox{qm}}(\beta)\;
\ .
\end{eqnarray}
Then $w_{\mbox{qm}}(\beta)$ is the thermal weight in this phase
space and ${\mathcal B}(r,p)=\bra{r,p}\op{B} \ket{r,p}$ a
function of the phase space variables.

\newpage
\section{Classical limit}

The connection to classical mechanics is established either by
performing the classical ($\hbar\rightarrow 0$) or the high
temperature limit ($\beta\rightarrow 0$).  The exponential in
the exponent of $w_{\mbox{qm}}(\beta)$ can be expanded,
$\hbar\omega$ drops out and the weight \fmref{E-7} approaches the
classical result
\begin{eqnarray}
\label{E-9}
w_{\mbox{cl}}(\beta)
&=&
e^{-\beta {\mathcal H}(z)}
\ .
\end{eqnarray}
Keeping the next order in the Taylor expansion of the exponential
function one gains
\begin{eqnarray}
\label{E-10}
w_{\mbox{no}}(\beta)
&=&
e^{-\beta \left(1+\half\beta\hbar\omega\right)
{\mathcal H}(z)}
\ .
\end{eqnarray}
\begin{figure}
\begin{center}
\epsfig{file=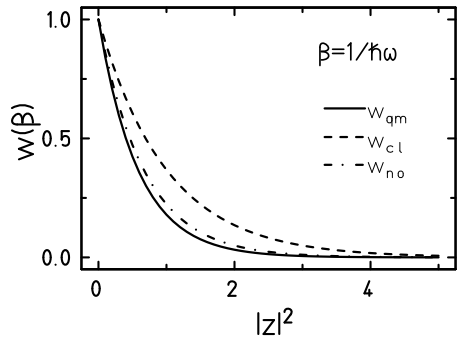,width=60mm}
\caption{Quantum weight (solid line, eq. \fmref{E-7}), classical
weight (dashed line, eq. \fmref{E-9}), next order (dashed
dotted line, eq. \fmref{E-10}) for the harmonic oscillator potential.}
\label{F-1}
\end{center}
\end{figure}
Figure \xref{F-1} shows all three weights for a specific inverse
temperature $\beta=1/(\hbar\omega)$. One sees that at the same
temperature the quantum mechanical distribution is much narrower
in $|z|^2\propto {\mathcal H}(z)$ than the classical one,
meaning that the quantum mean excitation energy is less than the
classical mean energy.

\section{Conclusions}

The purpose of the above exercise is to remind that one can
represent quantum mechanics and statistics in a basis which is
characterized by a continuous parameter. Although this
parameter {\it does not know anything about the discrete nature
of the energy eigenvalues} and moreover leads to the temptation
of a classical interpretation, the quantum statistical results
are correct, because they do not depend on the chosen
representation.

Therefore, if the time evolution of a particle enclosed in a
harmonic oscillator and described by a coherent state populates
the Hilbert space, or phase space in the sense of \fmref{E-8},
with the weight $w_{\mbox{qm}}(\beta)$ the statistical
properties are quantum mechanically correct.

It is a different question whether a specific approximation of
the time-dependent Schr\"odinger equation or a specific coupling
to a thermostat actually results in such a time evolution. 
In the spirit of Nos\'e \cite{Nos91} one could for example define a
thermostat which works by scaling coordinates and momenta so
that the coherent state visits the Hilbert space according to
the desired weight. One possible Nos\'e Hamiltonian would be
\begin{eqnarray}
{\mathcal H}(r,p,s,p_s)
&=&
\frac{p^2}{2 m s^2} + \half m \omega^2 s^2 r^2
+
\frac{p_s^2}{2 M} + \frac{\hbar\omega}{e^{\beta\hbar\omega}-1} \ln(s)
\ .
\end{eqnarray}
Another example, which is formulated with the help of a
thermostat for a constrained dynamical system, is given in
ref. \cite{Kus93}.

Concerning approximations of the time-dependent Schr\"odinger
equation it could be shown numerically that distinguishable
particles contained in a common harmonic oscillator field and
interacting via a short-ranged two-body force which are
described by coherent states approach classical statistics on
time-average \cite{ScF96}, but that the extension of the
coherent state towards wave packets with time-dependent width
(squeezed coherent states) as used in FMD \cite{FeS97}
\begin{eqnarray}
\braket{x}{r,p,a}
&\propto&
\exp\left\{-\frac{(\,{x}-r\,)^2}{2\,a}
+\frac{i}{\hbar}px\right\} 
\end{eqnarray}
is sufficient to obtain the quantum result.
\begin{figure}
\begin{center}
\epsfig{file=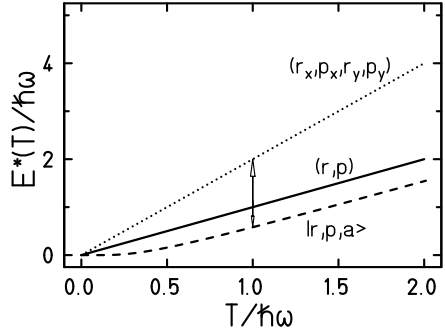,width=58mm}
\caption{Excitation energy for different degrees of freedom. The
solid line shows the classical result for a one-dimensional
harmonic oscillator, the dotted line the result for
two-dimensions; the dashed line displays what happens if the
coherent state is replaced by a wave packet with time-dependent width.}
\label{F-2}
\end{center}
\end{figure}
Having a closer look on the excitation energy
\begin{eqnarray}
{\mathcal H}(r,p,a)
&=&
\bra{r,p,a}\op{H} \ket{r,p,a}-\half\hbar\omega 
\\
&=&
\frac{p^2}{2 m} + \half m \omega^2 r^2
+
\frac{\hbar}{2}\left(
\frac{1}{2 m\,\Re{a}}+m \omega^2 \frac{|a|^2}{2 \,\Re{a}}
\right)
-\half\hbar\omega 
\nonumber
\end{eqnarray}
which now depends not only on the degrees of freedom $r$ and
$p$, but also on the width parameter $a$, this might seem
counter intuitive compared to classical statistics. As
illustrated in fig.~\xref{F-2} in classical statistics
additional degrees of freedom, for instance the extension of
the one-dimensional oscillator to two dimensions, lead to higher
mean energy at the same temperature, whereas the enlarged
freedom of the wave function introduced by the non-classical
width parameter does the opposite. The wave function is now able
to ``feel" the underlying density of states which is less than
the classical one, therefore, less excitation energy belongs to
the same temperature.

In general one must say that whether an approximate quantum
dynamics is able to reflect thermal averages does depend on the
freedom of the trial state and also on the observables one is
interested in. It may very well be that the parameterization of
the trial state is rich enough for single-particle observables
but not for many-particle observables.

\stars 

I would like to thank the National Institute for Nuclear Theory
at the University of Washington for the warm hospitality during
the program INT--98--2 where a part of this work was done and I
would like to thank the Deutsche Forschungsgemeinschaft (DFG) for
financial support of this visit.  I am thankful for stimulating
discussions to G.~Bertsch, A.~Bulgac, P.~Borrmann, H.~Feldmeier,
P.G.~Reinhard, R.~Roth, H.~Schanz and E.~Suraud and to
K.~B\"arwinkel and T.~Neff for carefully reading the manuscript.

%
%


\vskip-12pt
\begin{thebibliography}{99}
%
\bibitem{EbM97}
\Name{W. Ebeling, B. Militzer} 
\Review{Phys. Lett.} 
\Vol{A226}
\Year{1997} 
\Page{298}.
\bibitem{OhR97B}
\Name{A. Ohnishi, J. Randrup} 
\Review{Phys. Rev.} 
\Vol{A55}
\Year{1997} 
\Page{R3315}.
\bibitem{AiS86}
\Name{J. Aichelin, H. St\"ocker} 
\Review{Phys. Lett.} 
\Vol{B176}
\Year{1986} 
\Page{14}.
\bibitem{Fel90}
\Name{H. Feldmeier} 
\Review{Nucl. Phys.} 
\Vol{A515}
\Year{1990} 
\Page{147}.
\bibitem{OHM92}
\Name{A. Ono, H. Horiuchi, Toshiki Maruyama, A. Ohnishi}
\Review{Phys. Rev. Lett.}
\Vol{68}
\Year{1992}
\Page{2898}.
\bibitem{KlS85}
\Name{J.R. Klauder, B.-S. Skagerstam} 
\Book{Coherent States}
(World Scientific Publishing Co. Pte. Ltd., Singapore)
\Year{1985}.
\bibitem{ScF97}
\Name{J. Schnack, H. Feldmeier} 
\Review{Phys. Lett.} 
\Vol{B409}
\Year{1997} 
\Page{6}.
\bibitem{Nos84}
\Name{S. Nos\'e} 
\Review{J. Chem. Phys.} 
\Vol{81}
\Year{1984} 
\Page{511}.
\bibitem{Hoo85}
\Name{W.G. Hoover} 
\Review{Phys. Rev.} 
\Vol{A31}
\Year{1985}
\Page{1685}.
\bibitem{KBB90}
\Name{D. Kusnezov, A. Bulgac, W. Bauer} 
\Review{Ann. of Phys.} 
\Vol{204}
\Year{1990} 
\Page{155}.
\bibitem{Nos91}
\Name{S. Nos\'e} 
\Review{Prog. of Theor. Phys. Suppl.} 
\Vol{103}
\Year{191} 
\Page{1}.
\bibitem{Kus93}
\Name{D. Kusnezov} 
\Review{Phys. Lett.} 
\Vol{A184}
\Year{1993} 
\Page{50}.
\bibitem{OhR93}
\Name{A. Ohnishi, J. Randrup} 
\Review{Nucl. Phys.} 
\Vol{A565}
\Year{1993} 
\Page{474}.
\bibitem{KTR94A}
\Name{D. Klakow, C. Toepffer, P.-G.Reinhard} 
\Review{Phys. Lett.} 
\Vol{A192}
\Year{1994} 
\Page{55}.
\bibitem{BDH94}
\Name{P. Blaise, P. Durand, O. Henri-Rousseau} 
\Review{Physica} 
\Vol{A209}
\Year{1994} 
\Page{51}.
\bibitem{ScF96}
\Name{J.Schnack, H. Feldmeier} 
\Review{Nucl. Phys.} 
\Vol{A601}
\Year{1996} 
\Page{181}.
\bibitem{OnH96}
\Name{A. Ono, H. Horiuchi}
\Review{Phys. Rev.} 
\Vol{C53}
\Year{1996} 
\Page{2341}.
\bibitem{OhR97}
\Name{A. Ohnishi, J. Randrup} 
\Review{Phys. Lett.} 
\Vol{B394}
\Year{1997} 
\Page{260}.
\bibitem{Sch98}
\Name{J. Schnack} 
\Review{Physica} 
\Vol{A259}
\Year{1998} 
\Page{49}.
\bibitem{KTR94B}
\Name{D. Klakow, C. Toepffer, P.-G.Reinhard} 
\Review{J. Chem. Phys.} 
\Vol{101}
\Year{1994} 
\Page{1}.
\bibitem{FeS97}
\Name{H. Feldmeier, J. Schnack} 
\Review{Prog. Part. Nucl. Phys.} 
\Vol{39}
\Year{1997} 
\Page{393}.
\bibitem{INT98}
http://www.phys.washington.edu/$\sim$bulgac/int\_98.html
\bibitem{AEM95}
\Name{M.H. Anderson, J.R.~Ensher, M.R.~Matthews, C.E.~Wieman,
E.A. Cornell} 
\Review{Science} 
\Vol{269}
\Year{1995} 
\Page{198}.
\bibitem{DMA95}
\Name{K.B. Davis, M.-O. Mewes, M.R.~Andrews, N.J.~van Druten,
D.S.~Durfee, D.M.~Kurn, W.~Ketterle} 
\Review{Phys. Rev. Lett.} 
\Vol{75}
\Year{1995} 
\Page{3969}.
\bibitem{BSH97}
\Name{C.C.~Bradley, C.A.~Sackett, R.G.~Hulet} 
\Review{Phys. Rev. Lett.} 
\Vol{78}
\Year{1997} 
\Page{985}.
\bibitem{Sch96}
\Name{J. Schnack} 
\Book{dissertation} 
(TH Darmstadt)
\Year{1996}.
\end{thebibliography}
\end{document}